# Quantitative phase imaging via Fourier ptychographic microscopy

Xiaoze Ou,[1,†] Roarke Horstmeyer,[1,†] Changhuei Yang[1], and Guoan Zheng[1,2,*]

[1]*Electrical Engineering, California Institute of Technology, Pasadena, CA, 91125, USA*
[2]*Presently at Biomedical Engineering & Electrical Engineering, University of Connecticut, Storrs, CT, 06269, USA*
[†]*These authors contributed equally to this work*
*\*Corresponding author: guoan.zheng@uconn.edu*

Fourier ptychographic microscopy (FPM) is a recently developed imaging modality that uses angularly varying illumination to extend a system's performance beyond the limit defined by its optical elements. The FPM technique applies a novel phase retrieval procedure to achieve both resolution enhancement and complex image recovery. In this letter, we compare FPM data to both theoretical prediction and phase-shifting digital holography measurement to show that its acquired phase maps are quantitative and artifact-free. We additionally explore the relationship between the achievable spatial and optical thickness resolution offered by a reconstructed FPM phase image. We conclude by demonstrating both enhanced visualization and the collection of otherwise unobservable sample information using FPM's quantitative phase.

The challenge of recovering quantitative phase information from a specimen's digital image has stimulated the development of many computational techniques over the past several decades. Such techniques, collectively referred to as phase retrieval algorithms, have had significant impact in simplifying the complexity of phase measurement setups in optical [1], X-ray [2] and electron imaging [3] experiments.

The Gerchberg-Saxton (GS) algorithm [4] is one of the earliest strategies for recovering a specimen's phase from intensity measurements. In general, this iterative procedure alternatively constrains the specimen's complex solution to conform to the measured intensity data in the spatial domain, and to obey a known constraint in the Fourier domain. While proven to weakly converge, stagnation and local minima issues limit its applicability [5]. Gonsalves [6] and Fienup [5, 7] both recognized that applying multiple unique intensity measurement constraints, as opposed to a single intensity constraint, helps prevent stagnation and greatly improves convergence speed. This type of "phase diversity" procedure now includes variants based on translational diversity [8], defocus diversity [9], wavelength diversity [10, 11], and sub-aperture piston diversity [12].

Of particular interest to this letter are phase retrieval schemes based on translational-diversity (i.e., moving the sample laterally). A related technique termed ptychography [13-15], often applied with X-ray [16] and electron microscope imagery [17], can both acquire phase and improve an image's spatial resolution. While setups exist in many flavors [18-24], the general ptychographic approach consists of three major steps: 1) illuminating a sample with a spatially confined probe beam and capturing an image of its far-field diffraction pattern, 2) mechanically translating the sample to multiple unique spatial locations (i.e., applying translational diversity) while repeating step 1, and 3) using the set of captured images as constraints in an iterative algorithm. Details regarding ptychography's operation are in [14, 18], and demonstrations of its quantitative phase performance are in [17-24], which have also been extended to the optical regime [25-27]. It is important to note that ptychography achieves resolution improvement by physically scanning its probe over an extended field-of-view, and the computational acquisition of phase is vital to the accurate fusion of its acquired low-resolution imagery.

Recently, a unique implementation of ptychography in the Fourier domain, termed Fourier Ptychographic Microscopy (FPM) [28], was introduced to extend an optical imaging system's resolution. The goal of this paper is to prove how and why FPM can capture accurate quantitative phase measurements, which was not addressed in [28] at all. The FPM setup and a schematic of its algorithm are in Fig. 1. FPM uses no mechanical movement to image well beyond a microscope's traditional cutoff frequency. Unlike conventional ptychography, FPM uses a fixed array of LED's to illuminate the sample of interest from multiple angles. At each illumination angle, FPM records a low-resolution sample image through a low numerical aperture (NA) objective lens. The objective's NA imposes a well-defined constraint in the Fourier domain. This NA constraint is digitally panned across the Fourier space to reflect the angular variation of its illumination. FPM converges to a high-resolution complex sample solution by alternatively constraining its amplitude to match the acquired low-resolution image sequence, and its spectrum to match the panning Fourier constraint. As a combination of phase retrieval [5-12] and synthetic aperture microscopy [29-31], it is clear that phase must play a vital role in successful convergence.

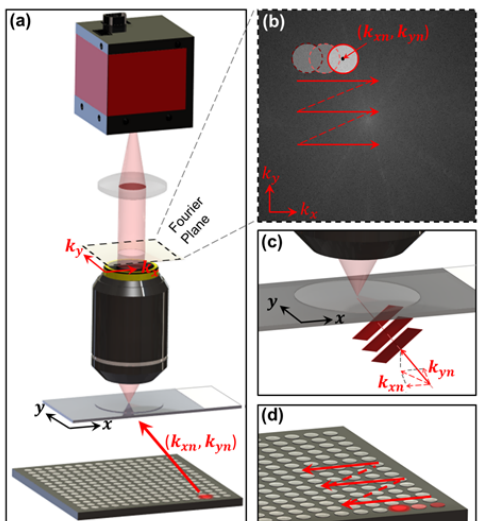

Fig. 1. FPM setup and imaging procedure. (a) An LED array sequentially illuminates the sample with different LED elements. (b) The object's finite spatial frequency support, defined by the microscope's NA in the Fourier domain (red circle), is imposed at offset locations to reflect each unique LED illumination angle. The Fourier transform of many shifted low-resolution measurements (each circle) are stitched together to extend the complex sample spectrum's resolution well beyond the objective lens's cutoff. (c) Light emitted from a single LED strikes a small sample area with wavevector $(k_{xi}, k_{yi})$. (d) LEDs are sequentially activated during FPM image acquisition.

While [28] demonstrated that FPM can accurately render improved-resolution intensity images, the accuracy of FPM phase remains in question. There is no guarantee that the phase acquired through FPM's iterative process must quantitatively match the sample – a multitude of possible phase distributions could allow its non-convex algorithm to map the acquired data set to an accurate high-resolution intensity image. One would additionally expect the limited spatial coherence of FPM's illumination to further compound any attempted complex field reconstruction. Finally, since much of the images' redundant information is utilized to improve spatial resolution, it is not clear if, and at what resolution, a simultaneously acquired phase map will deviate from ground truth. The primary goal of this paper is to prove that these challenges withstanding, FPM's phase images of thin samples are indeed quantitatively accurate, and thus deserve comparison with translation diversity and ptychography as an alternative "angular diversity" phase acquisition tool. Additional advancements include discussing this new system's phase resolution limits and demonstrating the acquired phase's ability to reveal additional information missing from intensity imagery. We intend the following work to cast FPM as a tool to accurately acquire not just intensity, but the full complex field produced by thin biological samples.

Our experimental system consists of a conventional microscope with a 15x15 red LED matrix (center wavelength 635 nm, 12 nm bandwidth, ~150 μm size) as the illumination source (Fig. 1). The 2D thin sample is inserted under a microscope's 2X, 0.08 NA objective lens. A sequence of 225 low-resolution intensity images are collected as the sample is successively illuminated by each of the 225 LEDs in the array. These images are input to FPM's phase retrieval algorithm that reconstructs a high-resolution map of the complex field at the sample plane. For example, the 500 × 500 pixel quantitative phase map in Fig. 3(a2) is generated from a sequence of 50 × 50 pixel cropped low-resolution images, an example of which is displayed in Fig. 3(a1).

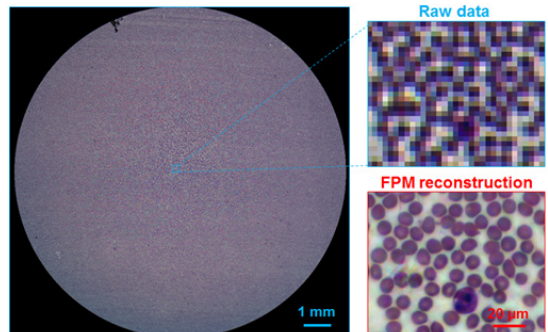

Fig. 2. Raw data and FPM intensity reconstruction of a blood smear. A 2X, 0.08 NA objective lens was used to capture the raw data. 225 low-resolution intensity images were used to recover the high-resolution FPM image.

This resolution gain is best understood by reviewing FPM's reconstruction algorithm. First, we initialize a high-resolution sample spectrum estimate $\hat{U}_0(k_x, k_y)$ as the Fourier transform of an up-sampled low-resolution image $I_{k_{xi},k_{yi}}(x,y) = I_{0,0}(x,y)$ captured under normal incidence. Second, this sample spectrum estimate is sequentially updated using the remaining 224 intensity measurements $I_{k_{xi},k_{yi}}(x,y)$, for i≠0, where subscript $(k_{xi}, k_{yi})$ corresponds to the illuminating plane wave's wavevector from the i[th] LED. For each update step, the sample spectrum estimate is shifted and multiplied by a known transfer function $T$: $\hat{U}_{i-1}(k_x - k_{xi}, k_y - k_{yi}) * T(k_x, k_y)$. The transfer function T is defined by shape of the back aperture of the microscope objective, typically a circle, as in Fig. 1(b). Next, a subset of this product is inverse Fourier transformed to the spatial domain to get $S_i$. The modulus of $S_i$ is then replaced by the square root of the known intensity $\sqrt{I_i}$ and transformed back to the spectral domain to create $\hat{S}_i$. Finally, the complex spectrum within the passband of the transfer function is replaced by the updated spectrum $\hat{S}_i$ to form a new sample spectrum estimate $\hat{U}_i$. The constraint-and-update sequence (identical to phase retrieval) is repeated for all i∈(1, 225) intensity measurements, as shown in Fig. 1(b). Third, we iterate through the above process several times until solution convergence, at which point $\hat{U}$ is transformed to the spatial domain to offer a high-resolution complex sample image.

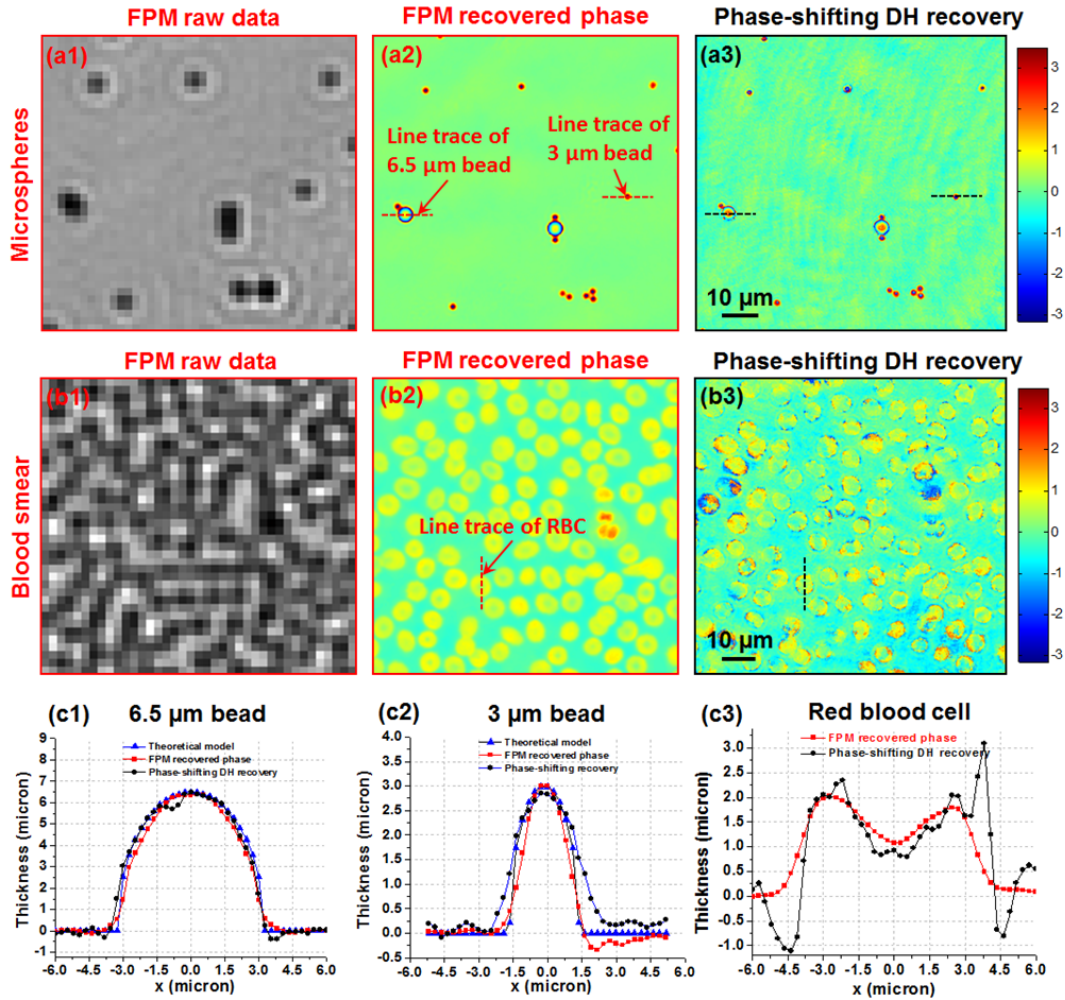

Fig. 3. Comparing FPM phase reconstructions to digital holographic and theoretical data. FPM transforms low-resolution intensity images from a 2X objective (a1) into a high-resolution phase map (a2) of different-sized polystyrene microbeads, as compared with a DH reconstruction (a3) using a 40X objective. (b) A similar image sequence highlights FPM's phase imaging capabilities on a human blood smear. (c) Line traces through the microbeads and a RBC demonstrate quantitative agreement with expected phase performance.

Fig. 2 demonstrates how the data acquisition and post-processing scheme outlined above can greatly improve the resolution of measured optical intensities. To verify FPM's ability to also accurately recover optical phase, we imaged a sample containing microbeads in oil (3 μm and 6.5 μm diameter, $n_{oil}$ = 1.48, $n_{sphere}$ = 1.6), shown in Fig. 3(a). Unwrapped line traces of the optical phase shift induced by two different-sized spheres lead to estimated microbead thickness curves in Fig. 3(c1)-(c2), exhibiting close agreement with theory. The root mean-squared error (RMSE) between experimental and theoretical thickness is 0.25 μm and 0.33 μm, respectively.

A phase-shifting digital holography (DH) microscope with a 40X objective lens also provides experimental ground-truth comparison. Our DH setup splits a solid-state 532 nm laser into a sample and reference arm (both spatially filtered and collimated). The reference arm passes through an electro-optic phase modulator (Thorlabs EO-PM-NR-C1) before recombination with the sample beam for imaging (Prosilica GX 1920, 4.54 μm pixels) via an objective (40×, 0.65 NA Nikon Plan N) and tube lens. 4 images are captured with a π/2 phase shift added to the reference between each image. Sample phase is calculated from the 4 images via the phase recovery equation [32]. A RMSE of 0.41 μm and 0.30 μm for the 3 μm and 6.5 μm line traces also offer close agreement between the DH experimental measurements and theory.

Fig. 3(b) presents an FPM reconstruction of a complex biological sample – a human blood smear immersed in oil, a common quantitative phase measurement target [33]. The FPM and ground-truth DH phase maps closely match, as exhibited by the phase trace through a red blood cell in Fig. 3(c3) (MSE = 0.58 μm). Sources of error for the FPM setup include the inclusion of slight aberrations by the objective lens, effects of a partially coherent illumination source, and the influence of noise within the iterative reconstruction scheme. The primary source of error in the DH data is speckle "noise" caused by a coherent illumination source. FPM phase tends towards a smoother phase profile in part because its LEDs' partially coherent illumination avoids coherent speckle artifacts.

A simple one-dimensional model helps describe limitations on the resolution of FPM's acquired phase image. From [28], we know FPM's maximum resolvable wavevector $k_x$ is limited by its maximum LED angle θ: $k_x^{max} = k(sinθ + NA)$. Likewise, the wavevectors emitted by a slowly varying phase object $\varphi(x)$ are set by its gradient: $k_x = d\varphi/dx$ in 1D. Assuming the phase object is a grating of period $p$ and thickness $t$, we can write $\varphi(x) = tsin(px)$. Using the above gradient relationship tells us its maximum emitted wavevector $k_x^{max} = tp$. Thus, the resolution limit for FPM phase is set by the product of the sample's spatial resolution and thickness, which both must be accounted for during system design. This argument extends to an arbitrary extended complex sample by Fourier-decomposing it into a finite set of gratings. While this relationship helped guide the design of the included experiments, a more detailed analysis is worth future investigation.

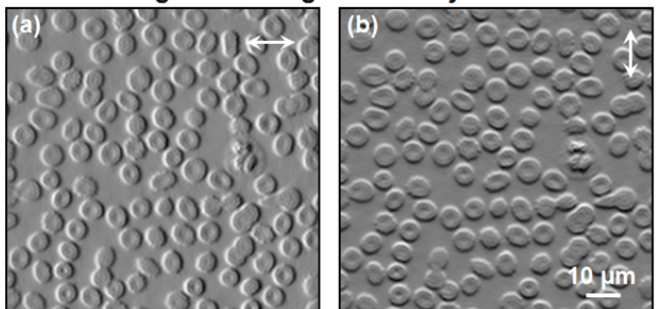

Fig. 4. Computed phase gradient images in x direction (a) and y direction (b) from the human blood smear phase map in Fig. 3.

The benefits of an acquired phase map are easily demonstrated with the computational generation of phase-gradient images in Fig. 4, simulating the improved visibility of a differential-interference-contrast microscopy. However, we note that this computational processing does not produce new information for the complex sample. Fig. 5 demonstrates how an acquired FPM phase map can give additional sample information otherwise absent from FPM's improved intensity resolution image.

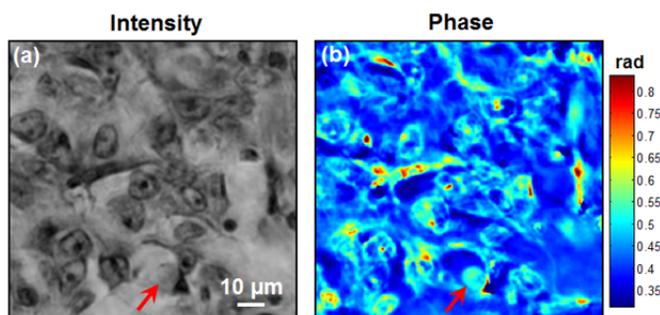

Fig. 5. FPM intensity and phase images of a tissue sample. As indicated by the red arrow, some cell feature is transparent in intensity image but visible in the phase image.

In conclusion, we have verified the FPM method can extract accurate and quantitative phase information from a set of raw intensity data, which may be useful for blood testing [34], tissue screening [35], and disease diagnosis [36]. We note that the accuracy of FPM reconstruction relies on sufficient spectrum overlapping in Fourier space. The relationship between data redundancy and the accuracy of reconstructed phase maps will be explored in detail in the future.

This work was supported by a grant from the National Institutes of Health (NIH) (NIH 1R01AI096226-01).